\documentclass[lettersize,journal]{IEEEtran}

\usepackage{graphicx}
\usepackage{url}
\usepackage{subfigure}
\usepackage{color}
\usepackage{booktabs}
\usepackage{fontawesome}
\usepackage{amsmath}

\usepackage{multicol} 
\usepackage{multirow} 
\usepackage[ruled]{algorithm2e}

\newcommand\fig{Fig.}

\begin{document}
\title{Investigating Inter-Satellite Link Spanning Patterns on Networking Performance in Mega-constellations}

\author{
	\IEEEauthorblockN{
		% Author1\IEEEauthorrefmark{1},
		Xiangtong~Wang\IEEEauthorrefmark{1},
		Xiaodong~Han\IEEEauthorrefmark{2},
		Menglong~Yang\IEEEauthorrefmark{1}\IEEEauthorrefmark{3}, 
		Chuan~Xing\IEEEauthorrefmark{2},
		Yuqi~Wang\IEEEauthorrefmark{2},
		Songchen~Han\IEEEauthorrefmark{1}, and
		Wei~Li\IEEEauthorrefmark{1}\IEEEauthorrefmark{3}
		}

			\IEEEauthorblockA{
		\IEEEauthorrefmark{1}School of Aeronautics and Astronautics, Sichuan University, Chengdu, China
	}

	\IEEEauthorblockA{
		\IEEEauthorrefmark{2}Robotic Satellite Key Laboratory of Sichuan Province, Chengdu, China
	}

	\IEEEauthorblockA{
		Email: \{li.wei, hansongchen\}@scu.edu.cn
	}
	\thanks{This paper was produced by the IEEE Publication Technology Group. They are in Piscataway, NJ.}% <-this % stops a space
\thanks{Manuscript received April 19, 2021; revised August 16, 2021.}

}
\markboth{Journal of \LaTeX\ Class Files,~Vol.~14, No.~8, August~2021}%
{Shell \MakeLowercase{\textit{et al.}}: A Sample Article Using IEEEtran.cls for IEEE Journals}

\maketitle

\begin{abstract}

    % 现有星座部署多数是为了优化覆盖或者碰撞避免，却鲜有考虑网络性能，例如容量，时延等因素，这在未来的卫星网络中具有重要意义。

    Low Earth orbit (LEO) mega-constellations rely on inter-satellite links (ISLs) to provide global connectivity. 
    We note that in addition to the general constellation parameters, the ISL spanning patterns are also greatly influence the final network structure and thus the network performance.
    
    In this work, we formulate the ISL spanning patterns, apply different patterns to mega-constellation and generate multiple structures.
     Then, we delve into the performance estimation of these networks, specifically evaluating network capacity, throughput, latency, and routing path stretch. 
The experimental findings provide insights into the optimal network structure under diverse conditions, showcasing superior performance when compared to alternative network configurations.

    \end{abstract}
    \begin{IEEEkeywords}
        Satellites networks, ISL pattern, network structure design, mega-constellation
        \end{IEEEkeywords}

      % ISL spanning pattern 根据卫星配置的链路数量可分为+Grid mode 和*Grid mode 两种，ISL pattern 和相位因子是影响网络结构的两个重要因素。

\IEEEpeerreviewmaketitle

\section{Introduction}

\IEEEPARstart{T}{he}  concept of a Low Earth Orbit (LEO) mega-constellation network gained significant attention in recent years. 
“NewSpace” companies are planning to launch hundreds to thousands of communication satellites into LEO in the future coming years.
 Their proposals have already obtained the regulatory approval: SpaceX\cite{starlink}, OneWeb\cite{oneweb}, and Telesat\cite{telesat} have acquired RF spectrum from the FCC for their constellations.
 The majority of these systems are configured in Walker\cite{walker1984satellite} configuration and are intended to be organized into a network that spanning both intra-orbit ISL (iISL) and inter-orbit ISL (or side links, sISLs) for providing low-latency global communication.

 While the exciting prospects outlined a blooming picture of the future integrated satellite-terrestrial networks, the community still lacks a comprehensive understanding of the topological characteristic and the network performance of modern mega-constellations.
\begin{figure}[t!]
    \begin{center}
        \includegraphics[width=0.9\linewidth]{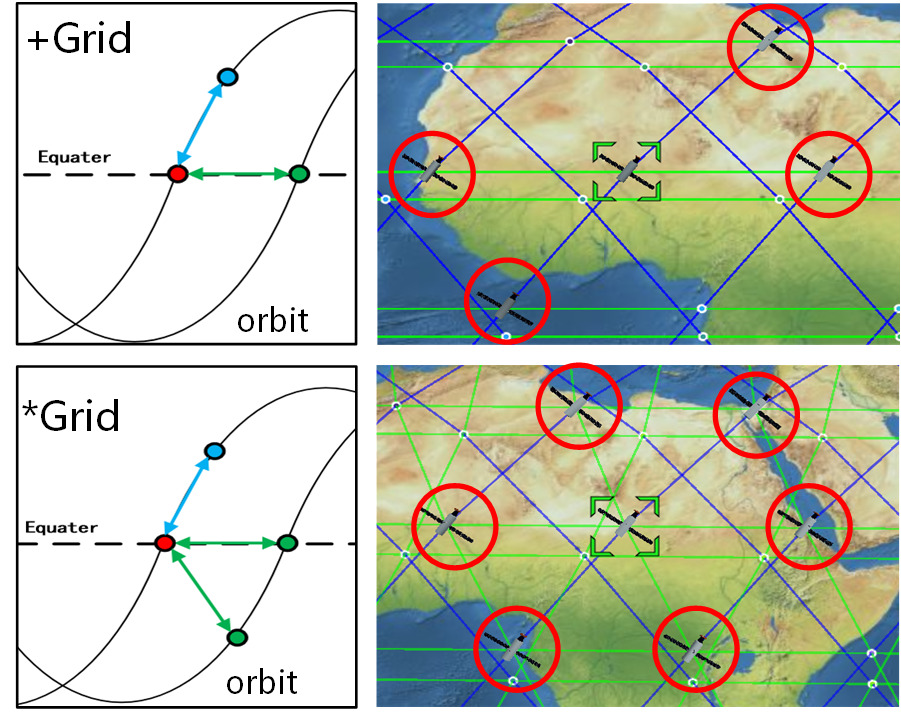}
    \end{center}
    \caption{Structure of a mega-constellation network formed by different ISL patterns.The first row is the structure of the `+Grid' mode, in which each satellite is adjacent to two sISLs, and the second row is the `*Grid' mode, in which each satellite is connected to four sISLs.} 
        \label{fig:teaser}
 \end{figure}
Several existing works\cite{chen2019topology,leyva2021interTcom,soret2019inter,yan2020integer} tried to model satellite matching problem and optimize the topology of novel constellations.
However, these efforts has primarily focused on the constellation with limited scale, such as Iridium-like network. These focus ignores the impact of different ISL spanning patterns on the global network or
the applicability to the emerging mega-constellations.
In references\cite{handley2018delay,bhattacherjee2019network,bhattacherjee2021towards,mclaughlin2023grid}, authors have analyzed the performance of global mega-constellation network, considering a certain proposed ISL connectivity mode such as `+Grid'\cite{bhattacherjee2019network} or `xGrid'\cite{mclaughlin2023grid}.
However, a conspicuous gap persists in the absence of a clear formulation regarding how the satellites connect to each other and an evaluation of network structures under various pattern configurations.

To reveal the structure characteristic of mega-constellation network formed by different ISL building configuration and its networking performance, we formulate the ISL spanning pattern and categorize the patterns into  `+Grid' and `*Grid' modes based on the number of ISLs per satellite.
We conduct evaluations under multiple constellations with varying ISL spanning patterns and satellite density. 
This allow us to validate the impact of structures formed by these patterns on the network performance.
Based on comprehensive experiments, we propose two optimal patterns for the `+Grid' and `*Grid' modes, respectively.
These patterns contribute to the best network performance in terms of multiple metrics.
To the best of our knowledge, this work is the first to formulate ISL spanning patterns and apply them to LEO Mega-constellations.
The structures of the network with different spanning patterns are shown in \fig\ref{fig:teaser}.

The contributions of this paper can be summarized as follows:

\begin{itemize}
    
    \item We formulate the ISL spanning patterns and apply them to mega-constellations, providing the visualizations of these structures;
    \item We evaluate the networks formed by various ISL spanning patterns using different metrics including path latency, stretch, capacity and throughput. Our evaluations enable us to identify the network structure with the optimal ISL spanning pattern.
\end{itemize}

\section{Model and Spanning Pattern Formulation}

% \input{tab/symbols.tex}

% \textbf{\textit{Definition 1:}}%time
\subsection{Network Model}

To address the temporal variations in the satellite networks,
we denote an ordered time set as $ \mathcal{T}= \{t_1, t_2,\cdots \}$.
% The elements of the set $\mathcal{T}$ are called time stamps, where $t_i \le t_{i+1}$.
The network topology without encountering ISLs\cite{eISL} can be considered unchanged between
adjacent time stamps. $t_{i+1}- t_i$ represents the minimum time granularity of scenario change.
% and the time granularity with each element represent one second of time can be simply described as a time set $\mathcal{T} = \{1, 2, T\}$. 
Therefore, the network topology at each time stamp
$t \in \mathcal{T}$ can be formulated as an undirected graph $\mathcal{G}^t = (\mathcal{V},\mathcal{E}^t )$, where $\mathcal{V}$ is the set of network vertices (satellites) and $\mathcal{E}^t$ is the set of undirected edges ( including iISL, sISLs and eISLs).% intro部分就列好ISL种类

The Walker\cite{walker1984satellite} constellation, which provide uniform coverage around the Earth, is generally described as $T/P/F/i$, where $T$ is the total number of satellites, $P$ is the number of equally spaced planes, $F$ is the phase factor and $i$ is orbit inclination. 
The change for satellites between neighboring planes is equal to $F \times 360 / T$.
The value range of the phase factor is $F \in [0, P-1]$. 
When $F=F_m$, the phase bias between satellites in two adjacent orbits achieves maximum where:
\begin{equation}
    \label{eq6}
   F_{m}=\left\{
    \begin{aligned}
    P/2 -1& , & P\text{~is~even.} \\
    (P-1)/2 & , &P\text{~is~odd.}
    \end{aligned}
    \right.
    \end{equation}
    % Note that the constellation formulates the same architecture as $F=0$ and $F=P$. 
% In this paper we use $N/P/F$ to illustrate a constellation where the $N$ is the number of satellites per orbit.

\subsection{ISL Spanning Pattern Formulation}

\label{sec:ISL}
\textbf{Intra-orbit Links (iISL).}
Without considering orbital perturbations, each satellite in the same orbit (plane) follows the same direction and velocity.
If a satellite establishes a link only with a neighbor satellite in the same orbit, the topology can be considered invariant. This is the only case of intra-orbit link in this paper.

\textbf{Inter-orbit links (side links, sISL).}
\label{sec:sISL}
Inter orbit links, i.e., the side links, refer to connections established between two satellites that move in the same direction but within side orbits.

\begin{figure}[htbp]
    \begin{center}
        \includegraphics[width=0.8\linewidth]{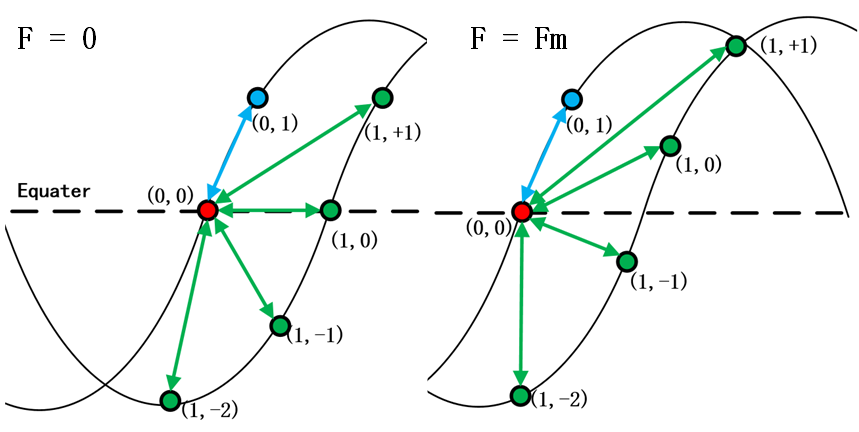}
    \end{center}
    \caption{Possible sISLs (green) of satellite (0,0) in networks with $F=0$ and $F=F_m$.} 
        \label{fig:makeISLs}
 \end{figure}
The possible sISL between a satellite (red point) and its side-orbit satellites (green point) are shown in \fig\ref{fig:makeISLs}, including both $F=0$ (left) and $F=F_m$ (right) cases.
The satellite is denoted as a tuple $(n,f)$, where $n$ is orbit-number and $f$ is phase-number.
We only consider establishing links with side orbit satellite having phase bias within \{-2,~-1,~0,~1\}.
Based on this criterion, we formulate the ISL spanning pattern as a phase bias set:
\begin{eqnarray}
    \mathcal{B} \subseteq \{-2,~-1,~0,~1\}
\end{eqnarray}

If each satellite is equipped with 4 ISLs, comprising 2 iISLs and 2 sISLs, the pattern corresponds to the `+Grid' mode, as described in \cite{bhattacherjee2019network}. 
In this configuration, the number of elements in phase bias set is $\left\lVert \mathcal{B} \right\rVert =1$.
Conversely, if each satellite has 6 ISLs, including 2 iISLs and 4 sISLs, the pattern is categorized as the `*Grid' mode, and $\left\lVert\mathcal{B}\right\rVert =2$.
Consequently, there are four distinct spanning patterns in the `+Grid' mode, represented by $\mathcal{B}=\{1\}$, $\mathcal{B}=\{0\}$, $\mathcal{B}=\{-1\}$, and $\mathcal{B}=\{-2\}$, with a more concise notation of \textit{b1, b0, bm1}, and \textit{bm2}, respectively.
Similarly, the `*Grid'mode encompasses six spanning patterns, which can be defined as $\mathcal{B}=\{1,0\}$, $\mathcal{B}=\{0,-1\}$, $\mathcal{B}=\{-1,-2\}$, $\mathcal{B}=\{1,-1\}$, $\mathcal{B}=\{1,-2\}$, and $\mathcal{B}=\{0,-2\}$, denoted as \textit{b10, b0m1, bm1m2, b1m1, b1m2} and \textit{b0m2}, respectively.
Consequently, the network configuration of these constellations is determined by the parameters $T/P/F/i$ and the phase bias set $\mathcal{B}$.
% the network structure is determined by both phase factor $F$ and ISL spanning pattern $\mathcal{B}$.
It's important to note that the number of transceivers on each satellite is constrained by cost considerations. Therefore scenarios involving satellites with more than four sISLs are not addressed in this paper.

 %   ISL pattern 与相位因子共同决定着网络结构。
% Note that there is another case where the ISL is established between two encountering satellites, which is denoted as encounter ISL (eISL).

%  For most LEO SNs, iISLs already provide sufficient SW-NE connectivity, while East-West connectivity and North-South connectivity require careful construction of sISLs.
%  \fig\ref{fig:ISLs} shows a constellation with the size $N =40, P=40$ and various configuration where all iISLs (left blue line) are identical, but sISLs (right green line) vary dramatically.
%  It can be seen that the more uniform the connectivity of the constellation in all directions, the better performance of networking overall if the user demand is evenly distributed globally.
%   More detailed analysis are shown in appendices.

\section{Network Evaluation Metrics}

\subsection{ISL direction distribution}

The phase factor $F$ and the bias set $\mathcal{B}$ play a crucial role in determining the ISL pattern, which inturn defines the global structure of the network.
This results in ISLs with different directional distributions.

It is known that, the sISLs in $F=0, \mathcal{B}=\{0\}$ pattern run parallel to the equator, which is called Horizontal Ring\cite{ekici2000datagram}.
This configuration effectively increases the connectivity of the network in the East-West direction\cite{handley2018delay}. 
In fact, directional connectivity reflects the distribution of ISLs over different directions, i.e., the higher the connectivity in a certain direction, the greater the proportion of ISLs within that angle, which will reduce the path zigzag in that direction and thus reduce the propagation latency.

In order to describe the directional connectivity of the constellation under different structures, we define the $\alpha$ as the angle between the ISL and Equator plane as:
\begin{eqnarray}
\alpha = \frac{\pi}{2} - \arccos{ (\frac{\mathbf{e}\cdot \mathbf{e}_z}{\|\mathbf{e}\|\cdot \|\mathbf{e}_z\|})}
\end{eqnarray}
where the $e_z$ is the normal vector of Equator plane, i.e., the rotation axis of Earth.
We count the probability density function $h(\alpha)$ during time $\mathcal{T}$ as follows: 
% which is shown in \fig\ref{fig:polar} at a polar coordination.
\begin{eqnarray}
    h^\mathcal{T}(\alpha) = \frac{\sum\limits_{\mathcal{T}}h^{t}(\alpha) }{\|\mathcal{T}\|}
    \label{eq:conn}
\end{eqnarray}

% 由于我们采用的链路模式包括+Grid和*Grid两种，所以还要统计*grid下，两种sISL组合后网络的连通性情况，因此提出通过互相关函数来对不同侧边链路的连通性进行衡量，即

% \begin{eqnarray}
%     R(l_1,l_2) = \int_{-\pi/2}^{\pi/2} h_{l_1}(\alpha) \cdot h_{l_2}(\alpha)
% \end{eqnarray}

% 若 R较大，则说明两种链路相关性大。为了得到更均匀的连通性，应选择相关性较小的两种拦链路组成*Grid pattern。

\subsection{Propagation Latency and Stretch}

% \noindent\paragraph{\textbf{Propagation Latency}}

For a specific network structure, we calculate the route between any two satellites within a period of time and derive the propagation latency:
\begin{eqnarray}
latency =  \frac{L_{prop}(p[s_i,s_j])}{c}, s_i,s_j \in S, 
\end{eqnarray} 
 where $L_{prop}(p[s_i,s_j])$ is the propagation distance of routing path that from $s_i$ to $s_j$, $S$ is the set of satellites and $c$ is light speed. 
%  The distribution of latencies can explain the performance differences of network structures.

% \noindent\paragraph{\textbf{Path Stretch}}

We define the stretch as the ratio of the path propagation distance $L_{prop}$ and the geodesic distance $L_{geo}$ between the same satellites pairs\cite{bhattacherjee2019network}, which is expressed as:
\begin{eqnarray}
    \label{eq:stretch}
stretch = L_{prop}/L_{geo}   
\end{eqnarray}

Given that the propagation speed of signal in an optical fiber is about $2c/3$, where $c$ is the speed of light, it can be inferred that if the $stretch \leq 1.5$, the path propagation latency in satellite network is lower than that in terrestrial fiber.

\subsection{Network Capacity and Throughput}

% \noindent\paragraph{\textbf{Network Capacity}}
% 网络容量与网络节点和网络链路有关，在理想情况下，当网络节点容量（即每秒处理数据量）和链路容量（每秒传输数据量）都达到最大，网络容量取二者中最小，即
Network capacity is contingent on the amount of data processed per second of satellites and data rate of ISLs in the networks. Under ideal situations, when both the capacity of satellites and ISLs (data rate) reach their maximum, the network capacity is determined by the lower of these two factors:
\begin{eqnarray}
    C(\mathcal{G}^t)  = \mathop{min}[ \sum\limits_{e \in \mathcal{E}^t} C(e), \sum\limits_{v \in \mathcal{V}} C(v) ]
\end{eqnarray}
% 在本文中，由于网络结构变化仅会影响链路容量， 这里我们不考虑卫星容量，并假设节点容量无限大，因此网络容量如下
where $e$ and $v$ represent the ISL and satellites, respectively.
Since changes in network structure primarily impact ISL capacity, we assume that satellite capacity is infinite. Therefore, the network capacity is $C(\mathcal{G}) = \sum\limits_{e_i \in \mathcal{E}} C(e_i)$.

% \noindent\paragraph{\textbf{Network Throughput}}

% 本文定义系统吞吐量量为某时刻网络多个连接対 在 路由算法下得到路径集合，其张成图的最大流
Network throughput is defined as the maximum flow of the graph spanned by the paths of multiple connections in the network at a given moment.
% 虽然图具有多个源点，多个汇点，但其最大流问题可以通过超级源点超级汇点来解决，
Although the graph has multiple source nodes and sink nodes, its maximum flow problem can be solved by introducing a super source and a super sink node\cite{west2001introduction}, and the throughput is calculated as follows:
\begin{eqnarray}
    \mathcal{T}(\mathcal{P}^t) = \mathop{maxflow}[ \mathcal{P}^t, v_{src},v_{dst}]
\end{eqnarray}
where $\mathcal{P}$ represents the set of routing paths, $ v_{src},v_{dst}$ is super source and source sink nodes, respectively. The link capacity of these super nodes to regular node is defined as infinity.
% 这里P为路经集合，v_{src} v_{dst}为超级源点汇点，其到普通点的链路容量定义为infinity
% {网络吞吐量}需要考虑负载，大多数情况下，吞吐量随着负载提升逐渐达到一个最优的定值，在网络的测量、协议的优化下，网络吞吐量应逐渐逼近容量。
Network throughput needs to consider the network loads and we define the loads as the number of connection paires. In most cases, throughput gradually reaches an optimal value as the load increases. 
% Through network measurement and protocol optimization, network throughput should gradually approach the capacity.

\section{Simulation Study}
% 本章，我们对不网络结构的性能的进行了探讨。
% 首先，我们说明+Grid 模式下，相同端点之间的路由路径区别以及其时延差异。
% 然后，我们对+Grid 和*Grid中 不同的网络结构，对网络的影响进行分析，并给出最优结构。
% 最后，我们对最优结构中不同数量卫星的星座性能进行分析。

In this section, we evaluate the satellite network performance affected by different network structures in terms of latency, path stretch, capacity and throughput.
We illustrate the differences and latency variations of routing path between two same end point (located at Harbin and London) under different network structure of `+Grid' mode.
Besides, we evaluate network performance under structures spanned by both `+Grid' and `*Grid' patterns, and the optimal structures are also given.
Furthermore, we evaluate network performance under different density constellations.

% the system throughput are given by calculating the capacity of each ISL.
% We finally show the end-to-end latency in random connection pairs distributed in two satellites, and the path stretch is also provided to illustrate the detour of routing paths.

% Our experiments were conducted using SNK\cite{snk}, a high-fidelity simulation system tool that enables satellite networking simulation and visualization.

\subsection{Path latency and stretch analysis}
%                         -----3.1 四种链路的方向性分析-------

\begin{figure}[htbp]
    \begin{center}
        \includegraphics[width=1\linewidth]{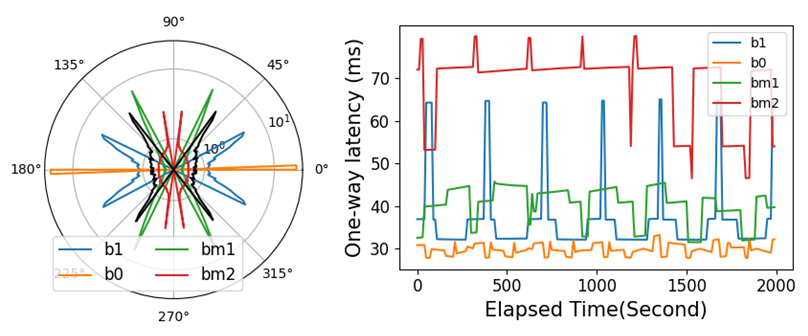}
    \end{center}
    \caption{ISL direction distribution (left) and path latency (right) in varying patterns.} 
        \label{fig:paths1}
 \end{figure}

% 图中所示为不同链路的方向分布，分布函数来自公式
% 可以看到iISL集中分布在53d， 这符合预期，因为其倾角为53°
%b1,b0,bm1,bm2链路分布差异较大，这会使得相同端节点之间的路由是不同的,例如东西方向终端的路由，在0°\180d链路分布更多的结构中，能取得更低的时延
The left of \fig\ref{fig:paths1} shows the probability density function (Eq.\ref{eq:conn}) of ISL direction. 
It illustrates that iISL is concentrated at $53^\circ$ (black line), which designs with expectations given the orbital inclination of $53^\circ$.
Patterns \textit{b1,b0,bm1,bm2} exhibit distinct distributions, which makes the routing between the same end nodes dramatically different. For example, the routing of east-west terminals, in the $0^\circ,~ 180^\circ$ link distribution more structure, can achieve lower latency.
% 如图1b是Harbin dao London之间的路由传播时延变化，路由算法使用dijstra
% 可以看到b0链路下的结构得到最低的时延，jitter最小，这是因为两个城市之间为东西分布，所有具有东西连通性最高的结构，即带有b0链路的网络，能让其得到最低的时延，
% 图2为同一时刻下路由路径，明显看到b0链路具有最短的路径。
The right of \fig\ref{fig:paths1} shows the one-way propagation latency between Harbin and London under Dijkstra routing algorithm.
The figure shows that compared to the ISL pattern of \textit{b1}, \textit{bm1}, and \textit{bm2}, the routing path achieves the lowest delay and jitter under the structure formed by the \textit{b0} pattern.

\begin{figure}[htbp]
    \begin{center}
        \includegraphics[width=1\linewidth]{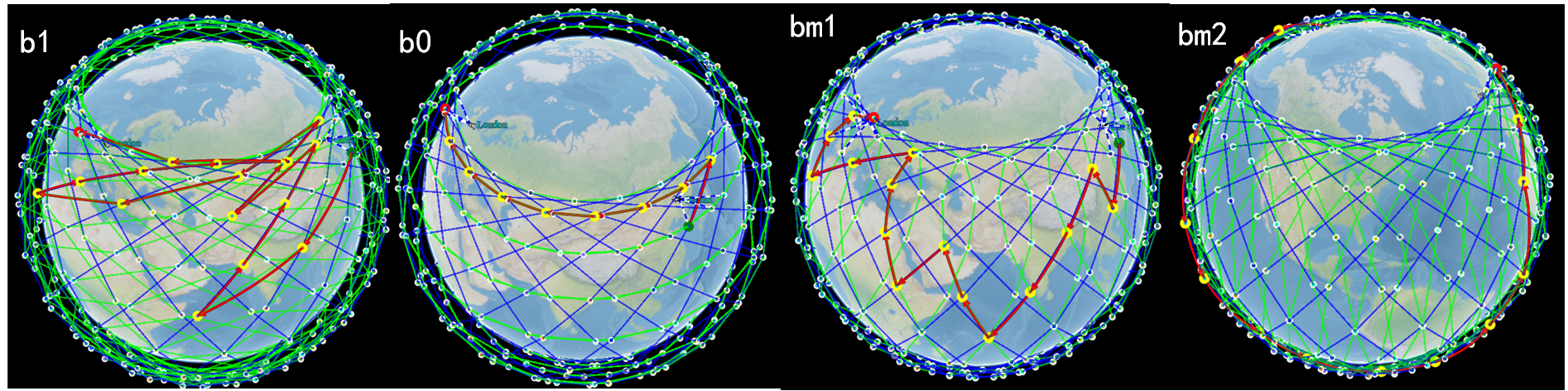}
    \end{center}
    \caption{Routing paths between Harbin and London in varying patterns (F=0).} 
        \label{fig:paths2}
 \end{figure}

This observation is attributed to the fact that London and Harbin share similar latitudes, resulting in ample connectivity from East to West.
The routing paths of above patterns are shown in \fig\ref{fig:paths2}.

% \begin{figure}[htbp]

%     \begin{center}
%         \subfigure[ISL direction distribution (left) and path latency (right) in varying patterns.]
%         {
%             \begin{minipage}{1\linewidth}
%                 \includegraphics[scale=0.423]{fig/lat4cities.png}%以这幅图的0.5倍大小输出
%             \end{minipage}
%         }
        
%         \subfigure[Routing paths between Harbin and London in varying patterns (F=0).]
%         {
%             \begin{minipage}{1\linewidth}
%                 \includegraphics[scale=0.3]{fig/paths.png}%以这幅图的0.5倍大小输出
%             \end{minipage}
%         }
%     \end{center}

%     \caption{The routing path and its latency between ends are both affected by ISL spanning patterns.}
%     \label{fig:conn}
% \end{figure}

%                         -----3.2 四种链路\六种组合的统计分析-------
% -----------latency and stretch-------------（这段有点难写）
%% outline：
%% 1. +*的对比；
%% vue:链路组合，cdf （链路组合的对比）
%% 结论，b0m1链路组合在f=0时具有最好的能力

% 为了更全面的统计出结构的差异，在任意时刻随机选择N个端点对，并通过dijstra路由计算出路径，并统计其时延和stretch分布情况。
% 图3a为

\begin{figure}[htbp]

    \begin{center}
        
    \subfigure[Latency in `+Grid' (left) and `*Grid' (right) patterns.]
    {
    \begin{minipage}{1\linewidth}
    \includegraphics[scale=0.36]{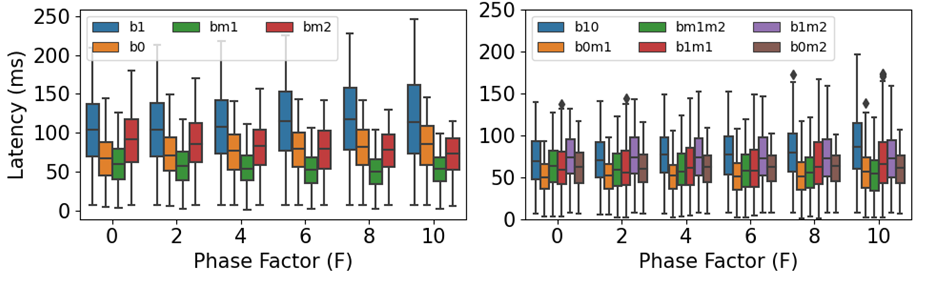}%以这幅图的0.5倍大小输出
    \end{minipage}}
	
	\subfigure[Stretch in `+Grid' (left) and `*Grid' (right) patterns.]
    {
    \begin{minipage}{1\linewidth}
    \includegraphics[scale=0.365]{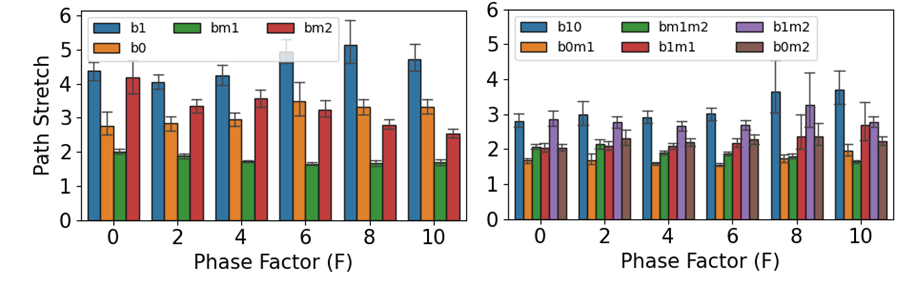}%以这幅图的0.5倍大小输出
    \end{minipage}
    }
\end{center}

    \caption{Latency and stretch distribution on networks with `+Grid' and `*Grid' mode patterns.} 
       \label{fig:latstre}
   \end{figure}

% 我们对不同相位因子下的四种sISL张成模式进行了分析
% 图中所示为不同链路模式下的 Dijstra路由算法下路径时延和伸张率情况，
% 可以看到：
% 平均时延bm1 < b0 < bm2 < b1
% 在b1,b0链路中，随着相位因子变大，平均时延逐渐增大；
% 在bm1 bm2链路中，随着相位因子增大，逐渐减小。
% 其中，在f=10, bm1的结构中，平均时延达到最低，超过80% 的路径时延不超过为80ms。
% 这说明，更均匀的连通性，能有效降低路径连接的绕路情况，以此降低时延。（转折）
% 我们对*Grid模式的结构，也进行了评估。图b可见，相比于+Grid模式明显减低了平均时延
% 其中，在相同相位下，b0m1模式时延最低，最高不超过70ms，平均在47ms
% 综上，+Grid结构中，bm1在F=Fm 下结构最优，而*Grid中b0m1结构最优
In our analysis of network structures defined by Inter-Satellite Link (ISL) patterns with varying phase factors, we noted that a more uniformly distributed connectivity proves to be more effective in minimizing the detour of routing path connections and, consequently, reducing latency.
\fig\ref{fig:latstre} (a) displays the distribution of paths latencies generated by Dijkstra under different network structures between random satellites.

The left of \fig\ref{fig:latstre} (a) shows that the networks with bm1 pattern has achieved the lowest average latency over all phase factors, following the trend $T_{bm1} \leq T_{bm0} \leq T_{bm2} \leq T_{bm1}$.
Within the structure of \textit{bm1, bm2} patterns, the latency is increases with the phase factor.
In contrast, in the structures spanned by \textit{b1, b0} patterns, the latency is decreases with the phase factor.
For the structure with bm1 pattern and $F=Fm$, the average latency is no more than 50ms and a maximum of no more than 80ms, which is only 40\% of the worst-case (bm1 pattern at $F=10$).
 The right of \fig\ref{fig:latstre} (a) shows the latency of `*Grid' structure that spanned by \textit{b10,~b0m1,~bm1m2,~b1m1,~b1m2} and \textit{b0m2} patterns.
 It is evident that the average latency is significantly reduced compared to the `+Grid' structure. In structures spanned by \textit{bm1m2,~b0m2} patterns, the latency decreases gradually with increasing phase factor while it remains relatively stable in \textit{b1m2} pattern.
In the structures spanned by \textit{b10, b0m1, b1m1} patterns, however, the latency increases with increasing phase factor, where the \textit{b0m1} mode has the lowest latency at the same phase. It reaches its lowest at $F=0$, with an average of 47ms and a maximum of 70ms.

We have also evaluated the path stretch that given by \ref{eq:stretch} under different structures, which is shown in \fig\ref{fig:latstre} (b).
Similarly, the structures in `+Grid' mode that spanned by bm1 patterns achieve the best results over all phase factors.
While in `*Grid' mode, the structure that spanned by \textit{b0m1} pattern with $F=0$ achieved the best result compared with the others.

\subsection{Capacity and throughput analysis}

\begin{figure}[htbp]
    \begin{center}
        \includegraphics[width=.9\linewidth]{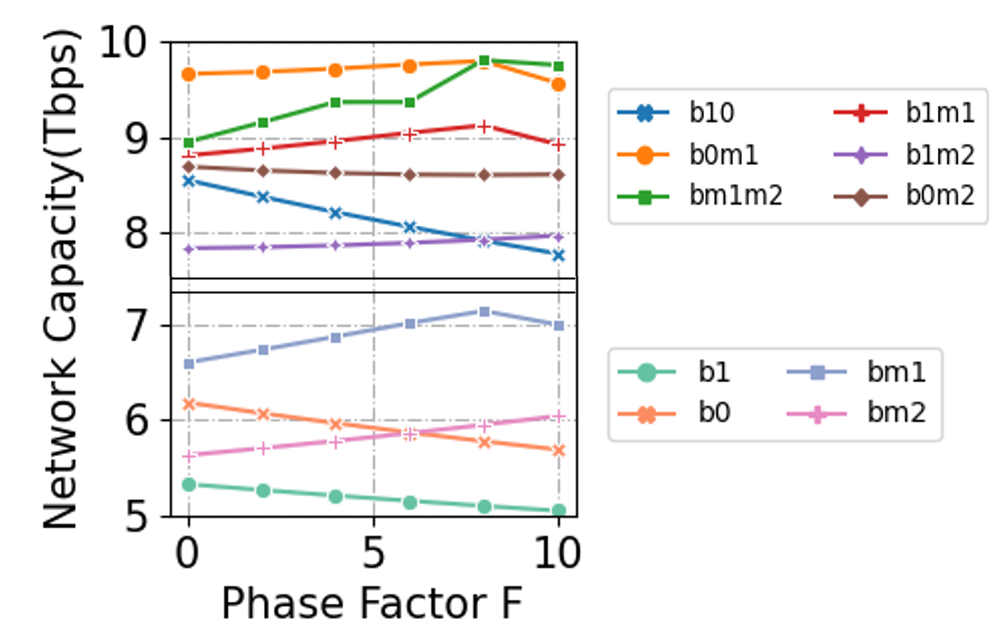}
    \end{center}
    \caption{Capacity on `+Grid' (left) and `*Grid' (right) network structures.} 
        \label{fig:cap}
 \end{figure}

% \begin{figure}[htbp]

%     \begin{center}
        
%     \subfigure[Capacity in +Grid structures.]
%     {
%     \begin{minipage}{.5\linewidth}
%     \includegraphics[scale=0.42]{fig/capG4.png}%以这幅图的0.5倍大小输出
%     \end{minipage}}\subfigure[Capacity in *Grid structures.]
%     {
%     \begin{minipage}{.48\linewidth}
%     \includegraphics[scale=0.42]{fig/capG6.png}%以这幅图的0.5倍大小输出
%     \end{minipage}
%     }
% \end{center}

%     \caption{Capacity on +Grid and *Grid network structures.} 
%        \label{fig:cap}
%    \end{figure}

% 由于星间链路张成模式不同，其链路容量也会由于不同的自由空间损耗而不同，从而影响网络容量不同。
% 图中所示为两种网络，多种结构在不同相位因子下容量的变化情况。

% 在+Grid中，b1 b0模式的结构随着F增加容量降低，而bm1 bm2则上升，
% BM1模式明显好于其他， 其中在F=8时，容量达到最大，7.2Tbps

% 在*Grid中，b10模式的结构随着F增加容量降低，而b1m1 bm1m2则上升，
% b0m1 b0m2,b1m2 则对相位因子的改变而变化缓慢，具有稳定性。
%其中 B0M1模式明显好于其他， 其中在F=8时，容量达到最大，9.75Tbps

%因此在大多数情况下，bm1链路和b0m1链路为最好结构，能达到最大的网络容量。
% 相比+Grid结构，*Grid结构容量在最好情况下增加了37%

The different ISL patterns result in varying link capacity due to different free space losses, thus affecting the network capacity differently.
The \fig\ref{fig:cap} illustrates how the network capacity changes for structures spanned by various ISL patterns.
On the right side of \fig\ref{fig:cap}, the capacity decreases as $F$ increases for the \textit{b1,~b0} patterns, while it increases for \textit{bm1} and \textit{bm2}. The bm1 pattern outperforms others, reaching its maximum capacity of 7.2~Tbps at $F=8$.
On the left side of \fig\ref{fig:cap} (b), the capacity decreases as $F$ increases for the \textit{b10} pattern, while it increases for \textit{b1m1} and \textit{b1m2}. The \textit{b0m1, b0m2}, and \textit{b1m2} patterns show slow variation with changes in the phase factor, demonstrating more stability features. The b0m1 pattern stands out as superior to others, reaching its maximum capacity of 9.75~Tbps at $F=8$.
Therefore, in most cases, \textit{bm1} links and \textit{b0m1} links represent the optimal structures, achieving the maximum network capacity. Compared to the `+Grid' configuration, the `*Grid' configuration shows a 37\% increase in capacity.

 \begin{figure}[htbp]
    \begin{center}
        \includegraphics[width=0.85\linewidth]{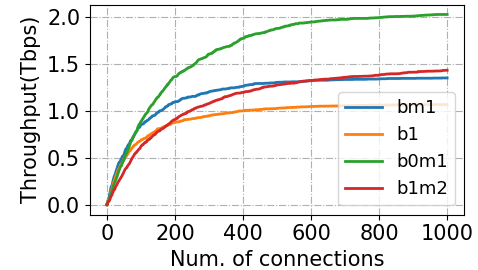}
    \end{center}
    \caption{Network throughput in structures with $F=0$.} 
        \label{fig:loads}
 \end{figure}

% 下图为bm1和b0m1结构下，吞吐量的变化。

% 不同的结构，除了能对容量造成影响，也会影响吞吐量。
% 我们选取了bm1,b1,b0m1,b1m2几个典型的结构，进行评估，其容量分别为+Grid 和*grid的最优和最差。
% 所用路由算法为dijstra，在每个时刻生成0-1000个连接对，作为负载的变量。
% 图中所示为不同结构网络在不同负载下的吞吐量情况。
% 在+Grid中，负载数量在400个连接时，b1 和bm1系统吞吐量已经收敛到1Tbps,1.35Tbps respectively, 吞吐量达到最大。
% 而在b1m2,b0m1结构中，负载数量在800左右时,吞吐量开始收敛，分别达到1.35Tbps and 2Tbps respectively,达极限.
% 可以看到，相同类别的结构，其吞吐量变化曲线变化率类似，即对负载数量反应一致，但更优的结构能得到更大的吞吐量，从而让网络具有更优的性能。在*Grid结构的吞吐量明显高于+Grid，这符合预期因为其具有更高的容量。

The throughput of different structured networks under different loads is shown in \fig\ref{fig:loads}.
We selected several typical structures, \textit{bm1,b1,b0m1,b1m2} for evaluation.
we generate 1000 random end-to-end pairs as the loads of network and calculate their paths under Dijkstra algorithm at each timestamp.
In patterns of `+Grid' mode, with a loads of 400 connections, the throughput of \textit{b1} and \textit{bm1} has converged to 1~Tbps and 1.35~Tbps, respectively.
In patterns of `*Grid' mode, the \textit{b1m2} and \textit{b0m1} structures, when the number of loads is around 800, the throughput starts to converge and reaches 1.35~Tbps and 2~Tbps, respectively.
We observe that the `+Grid' structures formed by \textit{bm1} patterns and `*Grid' structures formed by \textit{b0m1} offer superior  throughput while under the same routing scheme and loads.

In summary, \textit{bm1} is optimal pattern under $F=Fm$ in the `+Grid' structures, while \textit{b0m1} is optimal pattern in the `*Grid' structures.

\subsection{Density analysis}

\begin{figure}[htbp]

    \begin{center}

        \subfigure[Path stretch distribution.]
        {
            \begin{minipage}{0.45\linewidth}
                \includegraphics[scale=0.42]{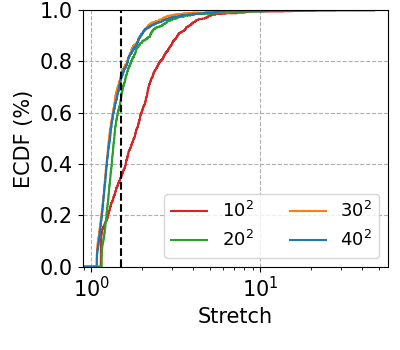}%以这幅图的0.5倍大小输出
            \end{minipage}
        }\subfigure[Path latency distribution.]
        {
            \begin{minipage}{.43\linewidth}
                \includegraphics[scale=0.42]{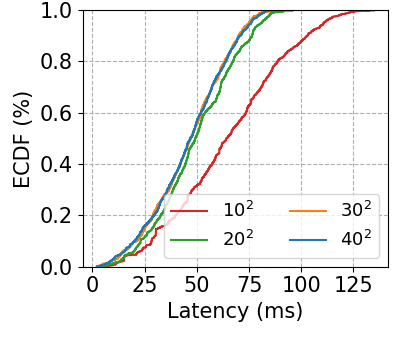}%以这幅图的0.5倍大小输出
            \end{minipage}
        }
    \end{center}

    \caption{ECDF of path stretch and latency in network structures with density $10^2,~20^2,~30^2,~40^2$.}
    \label{fig:cdf}
\end{figure}

Constellation networks of different density exhibit varying performance, where higher density satellites tend to have higher throughput and lower latency.
\fig\ref{fig:cdf} depicts the distribution of path stretch and latency in structures formed by \textit{b0m1} pattern at $F=0$ with densities $10^2$, $20^2$, $30^2$ and $40^2$, respectively.

In \fig\ref{fig:cdf} (a), the network with $10^2$ density exhibits the worst results, with more than 80\% of the paths having a stretch value of 2.9. In contrast, the other networks the value of about 1.6.
Additionally, over 70\% of the paths stretch in the networks of $20^2$, $30^2$ and $40^2$ is lower than 1.5, surpassing the performance of geodesic fiber transmission (black dash line).
In \fig\ref{fig:cdf} (b), we observe that the maximum latency in network with $10^2$ density exceeds 120ms, whereas it remains below 80ms in networks with density ~$20^2$ $30^2$ and $40^2$.
It illustrates that compared to the $10^2$ network, the $20^2$ network reduces the average latency from 75ms to 50ms, marking approximately a 33\% reduction. However, as the constellation size continues to grow, there is very limited improvement in the path latency and stretch, since the routing paths are getting approximate to the geodesic arc. Therefore, a network density of $20^2$ is sufficiently in the terms of path latency or stretch.

%                         -----3.3 *Grid中的尺度的统计分析-------
% 不同规模的星座网络性能不同，
% 图中为四种规模星座在*Grid，b0m1链路模式下的平均时延以及伸张率统计
% 图3a可以看到，随着星座尺寸变大，平均时延明显降低，在40的尺度下，平均时延达到30ms，相比10的尺寸，降低了87%

\vspace*{1pt}

\begin{figure}[htbp]
    \begin{center}
        \subfigure[Network throughputs.]
        {
            \begin{minipage}{0.55\linewidth}
                \includegraphics[scale=0.47]{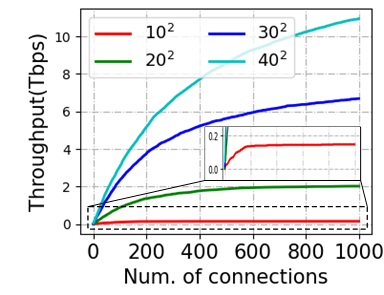}%以这幅图的0.5倍大小输出
            \end{minipage}
        }\subfigure[Network capacity.]
        {
            \begin{minipage}{.4\linewidth}
                \includegraphics[scale=0.47]{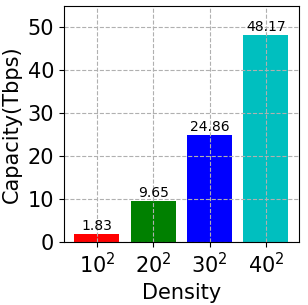}%以这幅图的0.5倍大小输出
            \end{minipage}
        }
    \end{center}

    \caption{Network capacity and throughput in structures with density $10^2,~20^2,~30^2,~40^2$.}
    \label{fig:thp_scale}
\end{figure}

We also provide the capacity and throughput among networks with different density.
\fig\ref{fig:thp_scale} shows the throughput and capacity of networks with different densities under `*Grid' mode spanned by \textit{b0m1} ISL pattern at $F=0$.
It illustrates the network with $10^2$ density reaches the throughput maximum at 200 connections, achieving about 0.16 Tbps. In contrast, the $20^2$ network reaches the maximum at 600 connections. $30^2$ and $40^2$ networks do not reach the maximum at more than 1,000 connections.
The theoretical capacity of $10^2$, $20^2$, $30^2$ and $40^2$ densities are 1.83~Tbps, 9.65~Tbps, 24.86~Tbps and 48.17~Tbps, respectively.

% 
% 图a中，10^2密度的网络在200个连接时达到了吞吐量最大值，大约0.16Tbps，而20^2 的网在600个连接够达到最大值。30^2 和40^2 的网络在超过1000个连接时没达到最大值。
% 10^2,20,30,40四种密度的理论容量分别为1。83,9.65,24.86 and 48.17Tbps
% 这说明随着网络规模的提升，容量与吞吐量都在明显提升，另外，由于路由算法仅使用基础的dijstra，没有负载均衡机制，容量利用率较低，约25%左右

\section{Conclusion}

% 本文主要探讨空间网络最优结构问题。
% 通过对多种结构的 吞吐量，路径伸张率、路径时延统计，我们得出结论：
% 1 若使用+Grid模式，即仅有两条sISLs，bm1模式性能最好，应将星座相位因子提升到最大，以此得到最好的平均时延和路劲伸张率；
% 2 若使用*Grid模式，即有四条sISLs，b0m1模式性能最好，并应将星座相位因子尽可能低，以此得到最低的时延和路径伸张率。
% 3 提升星座密度能有效提升网络性能，但对于时延提升有限,密度超过20^20意义不大，而更重要的是网络容量和吞吐量的提升.

This study addresses the challenges of determining optimal structure of LEO mega-constellation networks.
We formulate the ISL spanning patterns, apply different patterns to mega-constellation and generate multiple structures.
Through comprehensive experiments, we draw several conclusions:
1) In the case of utilizing the `+Grid' mode, it is observed that the \textit{bm1} pattern offers the most optimal structure. To achieve the best latency or throughput, it is advisable to maximize the constellation phase factor.
2) In the case of utilizing the `*Grid' mode, it is observed that the \textit{b0m1} pattern offers the most optimal structure. To achieve the latency or throughput, it is recommended to minimum the constellation phase factor.
3) Increasing the density of constellations can effectively improve the performance of the network, but the increase in latency is limited, and the density of more than $20^2$ is of little significance, while the more important thing is to improve the network capacity and throughput.

% \section{ACKNOWLEDGEMENT}
% This work has been supported by the 173 key project (no.2019-JCJQ-ZD-342-00).


\bibliographystyle{IEEEtran}
\bibliography{ref.bib}
\end{document}